\documentclass[12pt]{article}
\usepackage{amsmath,amssymb}
\usepackage{cite}
\usepackage{graphicx,color}
\usepackage{setspace}
\usepackage{nicefrac}
\usepackage[nottoc]{tocbibind} 
\usepackage{multirow,tabularx}

\usepackage[linktoc=all,hidelinks]{hyperref}

\usepackage{geometry}
\def\p{\partial}
\def\m{\mu}
\def\n{\nu}

\def\a{\alpha}
\def\b{\beta}
\def\d{\delta}

\def\e{\eta}

\def\f{\phi}

\def\k{\kappa}

\def\L{\Lambda}
\def\pr{\prime}

\def\r{\rho}

\def\t{\theta}
\def\T{\Theta}
\def\vf{\varphi}
\def\z{\zeta}
\def\nn{\nonumber}

\def\dif{\mathrm{d}}
\def\goesto{\rightarrow}

\def\cL{\mathcal{L}}

\def\cO{\mathcal{O}}

\def\tT{\widetilde{T}}

\newcommand{\link}[1]{[\href{http://arxiv.org/abs/#1}{{\tt arXiv:#1}}]}
\newcommand{\linkth}[1]{[\href{http://arxiv.org/abs/hep-th/#1}{{\tt arXiv/hep-th:#1}}]}

\newcommand{\doi}[1]{[\href{http://doi.org/#1}{{\tt doi:#1}}]}
\newcommand{\mail}[1]{\href{mailto:#1}{{\tt #1}}}

\numberwithin{equation}{section}

\usepackage{titletoc}
\dottedcontents{section}[1.2em]{}{1em}{0.5pc}
\dottedcontents{subsection}[4em]{}{2em}{1pc}

\begin{document}
\begin{titlepage}       


\begin{center}
	\setstretch{2}
	{\LARGE \bf Generalized Black Holes\\ in 3D Kerr-Schild Double Copy}
\end{center}

\begin{center}
\vspace{10pt}

{G{\"o}khan Alka\c{c}$\,{}^{a}$, Mehmet Kemal G\"{u}m\"{u}\c{s}$\,{}^{b}$ and Mehmet Ali Olpak$\,{}^{c}$}
\\[4mm]

{\small 
{\it ${}^a$No Affiliation}\\[2mm]

{\it ${}^b$Department of Physics, Faculty of Arts and Sciences,\\
	Middle East Technical University, 06800, Ankara, Turkey}\\[2mm]

{\it ${}^c$Department of Electrical and Electronics Engineering, Faculty of Engineering,\\ University of Turkish Aeronautical Association, 06790, Ankara, Turkey}\\[2mm]
e-mail: {\mail{gkhnalkac@gmail.com}, \mail{kemal.gumus@metu.edu.tr}, \mail{maolpak@thk.edu.tr }}
}
\vspace{5pt}
\end{center}

\centerline{{\large \bf{Abstract}}}
\vspace*{5mm}
\noindent The double copy of the Coulomb solution in three dimensions is a non-vacuum solution that can be obtained through different matter couplings. It is the static black hole solution of Einstein-Maxwell theory or general relativity minimally coupled to a free scalar field (with one ghost sign in the action in both cases). We consider generalizations of these matter couplings by paying particular attention to  the regularity of the static black solution on the gravity side and the corresponding single copy electric field in the gauge theory. We show that i) Einstein-Born-Infeld theory yields a singular double copy, which admits stable orbits for certain choices of parameters, with a regular single copy electric field. ii) Black hole solutions constructed in \link{2104.10172} by coupling to the scalar field exemplify mostly regular double copies with regular single copy electric fields and also admit stable orbits. Additionally, we use these solutions to investigate the connection between horizons on the gravity side and electric fields on the gauge theory side, which was previously observed in four dimensions.
\tableofcontents

\end{titlepage}

\section{Introduction}	
The double copy has become an indispensable tool in the study of scattering amplitudes since it allows one to obtain gravitational amplitudes by ``squaring" the Yang-Mills (YM) ones \cite{Bern:2010ue,Bern:2010yg}. It is also possible to construct perturbative classical solutions from semi-classical tree-level amplitudes \cite{Luna:2016hge,Goldberger:2016iau,Goldberger:2017frp,Goldberger:2017vcg,Goldberger:2017ogt,Shen:2018ebu,Carrillo-Gonzalez:2018pjk,Plefka:2018dpa,Plefka:2019hmz,Goldberger:2019xef,PV:2019uuv,Anastasiou:2014qba,Borsten:2015pla,Anastasiou:2016csv,Cardoso:2016ngt,Borsten:2017jpt,Anastasiou:2017taf,Anastasiou:2018rdx,LopesCardoso:2018xes,Luna:2020adi,Borsten:2020xbt,Borsten:2020zgj,Luna:2017dtq, Kosower:2018adc, Maybee:2019jus, Bautista:2019evw, Bautista:2019tdr, Cheung:2018wkq, Bern:2019crd, Bern:2019nnu, Bern:2020buy,Kosmopoulos:2021zoq, Kalin:2019rwq, Kalin:2020mvi, Almeida:2020mrg, Godazgar:2020zbv,Easson:2021asd,Chacon:2020fmr}. Although it seems impossible to establish a map between exact classical solutions due to the nonlinearity of both Einstein and Yang-Mills equations, one can obtain a linear structure for certain classes of spacetimes on the gravity side, and then, map them to the linearized solutions of YM theory, i.e., solutions of Maxwell's theory. The classical double copy, a non-perturbative manifestation of the double copy idea, can be realized via two different but related procedures. In the Kerr-Schild (KS) double copy \cite{Monteiro:2014cda}, one takes advantage of the fact that the Ricci tensor with mixed indices is linear in the perturbation for spacetimes admitting KS coordinates \cite{Gurses:1975vu}. Alternatively, one can work with the Weyl spinor (the spinor form of the Weyl tensor), which is a scalar under coordinate transformations but transforms under local change of the spinor basis. Analogous to the KS construction, for certain algebraically special spacetimes, one can find a spinor basis which linearizes the Weyl spinor and then relates it to a field strength spinor corresponding to a solution of Maxwell's equations. Although the Weyl double copy \cite{Luna:2018dpt} is more general, the two procedures give identical results, as they should, when both are applicable. There is a growing literature on the subject providing various examples and generalities \cite{Luna:2015paa,Bahjat-Abbas:2017htu,Carrillo-Gonzalez:2017iyj,Alkac:2021bav,Luna:2016due,Berman:2018hwd,Bah:2019sda,Banerjee:2019saj,Ilderton:2018lsf,Lee:2018gxc,Cho:2019ype,Lescano:2020nve,Lescano:2021ooe,Kim:2019jwm,Keeler:2020rcv,Huang:2019cja,Alawadhi:2019urr,Alawadhi:2020jrv,Elor:2020nqe,Berman:2020xvs,Easson:2020esh,Pasarin:2020qoa,CarrilloGonzalez:2019gof,Gumus:2020hbb,Alkac:2021seh}  (see \cite{Bern:2019prr,Kosower:2022yvp,Adamo:2022dcm} for reviews), and we now have a pretty good understanding of its origins and limitations thanks to results from twistor theory \cite{White:2020sfn,Chacon:2021wbr,Farnsworth:2021wvs}.

It has been known since the seminal work of Born and Infeld \cite{BI} that the singularity of the point charge solution at the origin might be removed by considering nonlinear generalizations of Maxwell's theory. It is natural to ask what the regularity of the single copy implies on the gravity side. In \cite{Pasarin:2020qoa}, by assuming that the standard KS ansatz for the double copy solution remains intact, the authors show that in order to ensure the regularity on the gravity side, the finiteness of the single copy electric field at the origin is not enough and it should vanish. For example, the double copy metric corresponding to the point charge solution of Born-Infeld electrodynamics is not regular. Recently, regular black hole metrics were constructed in \cite{Mkrtchyan:2022ulc} by starting from solutions of various non-linear electrodynamics theories with an electric field vanishing at the origin and employing the KS ansatz. Although this approach sheds light on the nature of the double copy solution, it is not clear how one should modify the fields equations on the gravity side such that the constructed metric is a solution. Therefore, it is desirable to study the regularity in the usual setup where one starts from a solution of a gravitational theory and maps it to a solution of Maxwel's theory. In 4D, such examples are given in \cite{Easson:2020esh}, and additionally, a simple but interesting connection between horizons on the gravity side and the zeros of the electric field on the gauge theory side was presented.

In this paper, we focus on the KS double copy in 3D\footnote{For recent developments in 3D double copy, see \cite{Moynihan:2020ejh,Burger:2021wss,Gonzalez:2021bes,Emond:2022uaf,Gonzalez:2021ztm,Hang:2021oso,Gonzalez:2022otg}}, which sets the stage for an important consistency check because there is no Newtonian limit of general relativity (GR) in 3D (see \cite{CarrilloGonzalez:2019gof} for a detailed explanation of why it might be, in principle, problematic for the KS construction). It was understood that the KS double copy works also in 3D with a crucial difference compared to higher dimensional versions: The double copy of the Coulomb solution is a non-vacuum solution \cite{CarrilloGonzalez:2019gof}. As a result of the on-shell duality of a scalar and a gauge vector in 3D, a static black hole with the required Newtonian potential ($\Phi \propto \log r$) that can be obtained by minimally coupling GR to Maxwell's theory  or a free scalar field provided that a ghost sign is taken for the Einstein-Hilbert or the matter term in the action \cite{Alkac:2021seh}. The former has the advantage that the matter field vanishes at infinity. In Einstein-Maxwell theory, one has the Coulomb electric field ($f_{rt} \propto \frac{1}{r}$). When the scalar coupling is used, the scalar field should be linear in the azimuthal angle ($\vf \propto \t$) to support the black hole solution. Our aim in this work is to consider generalizations of these matter couplings to extend the observations mentioned above to 3D. We will obtain static solutions admitting stable orbits, investigate the regularity of the single and double copy solutions, and uncover the relation between the horizons and the zeros of the electric field whenever possible.

The outline of this paper is as follows: In Section \ref{sec:KS}, we review the basics of the KS double copy for a static spacetime and discuss the regularity of the solutions. After an investigation of Einstein-Born-Infeld theory as a generalization of Einstein-Maxwell theory in Section \ref{sec:EBI}, we move on to generalizations of the scalar coupling in Section \ref{sec:sca}. We end our paper with conclusions and discussions in Section \ref{sec:sum}.

\section{Basics of the 3D Kerr-Schild Double Copy and Regularity of the Solutions}\label{sec:KS}
For a spacetime admitting KS coordinates, it is possible to write down the components of metric tensor in the following form \cite{Kerr}:
\begin{equation}
	g_{\m\n}=\e_{\m\n} + \phi k_\m k_\n\,,\label{KS}
\end{equation}
where $\phi$ is a scalar field and the vector $k_\m$ is null and geodesic with respect to the full metric $g_{\m\n}$ and the flat background metric $\e_{\m\n}$ (see chap. 32 of \cite{Stephani:2003tm} for a summary of important properties). In these coordinates, the Ricci tensor with mixed indices becomes linear in the perturbation as follows: 
\begin{equation}
	R^{\m}_{\ \n} =  \frac{1}{2}\left[\p^\a \p^\m (\phi k_\n k_\a)+\p^\a \p_\n (\phi k^\m k_\a)-\p^\a \p_\a (\phi k^\m k_\n) \right]\,. 
\end{equation} 
If one writes down the background line element in polar coordinates 
\begin{equation}
	\e_{\m\n}\dif x^\m\dif x^\n = -\dif t^2 + \dif r^2 + r^2 \dif \theta^2\,,
\end{equation}
and parametrizes the null vector as
\begin{equation}
	-k_\m \dif x^\m = \dif t + \dif r\,,
\end{equation}
the $\m0$-components become
\begin{equation}
	R^\m_{\ 0} = \frac{1}{2} \p_\n F^{\n\m}\,,\quad F_{\m\n}=2\p_{\left[\m\right.} A_{\left.\n \right]}\,,\quad A_\m \equiv \phi k_\m
\end{equation}
With minimal matter coupling, the trace-reversed gravitational field equations are given by
\begin{equation}
	R^\m_{\ \n} = \frac{\k^2}{2} \left(T^\m_{\ \n} - \d^\m_{\ \n} T\right)\,\qquad \k^2 = 8\pi G\,,
\end{equation}  
where $T_{\m\n}$ is  the energy-momentum tensor and $T$ is its trace. Checking the $\m0$-components, one obtains the Maxwell equations
\begin{equation}
	\p_\n F^{\n\m}=g J^\m
\end{equation}  
where the source is given by
\begin{equation}
	J^\m = 4 \left(T^\m_{\ 0} - \d^\m_{\ 0}  T\right)\,,
\end{equation}
and the gauge coupling is obtained by the identification\footnote{We choose our conventions such that when $G=1$, which is used in our numerical calculations, one has $\p_{\n}F^{\n\m}=2\pi J^\m$} $\k^2\rightarrow 4g$. Therefore, for each solution of the gravitational field equations that admit KS coordinates \eqref{KS}, the double copy, one can obtain a single copy solution of Maxwell's equations.

Unlike higher dimensions, in order to obtain the Coulomb's solution as the single copy, one needs matter coupling in 3D. One possibility is Einstein-Maxwell theory described by the action
\begin{equation}
	S= \int \dif^3 x\sqrt{-g}\left[\frac{1}{\k^2} R + \frac{1}{8\pi} f_{\m\n} f^{\m\n}   \right]\,,
\end{equation}     
where $f_{\m\n}=2\p_{\left[\m\right.} a_{\left.\n \right]}$. As shown in \cite{Alkac:2021seh}, one needs to introduce the Maxwell term with a ghost sign in order to obtain the correct Newtonian limit [$\phi=c\,\log(r)\,,c>0$]\footnote{One can also take the Einstein-Hilbert term with a ghost sign but throughout this paper, we will use the matter terms with a ghost sign.}. Taking $\phi = \phi(r)$, $a_\m \dif x^\m = a_t(r)\dif t$, one obtains the following static solution 
\begin{equation}\label{maxsol}:
	\phi(r)= - 8 G M - 2 G q^2 \log(r)\,,\quad a_\m \dif x^\m = -q \log(r) \dif t\,,
\end{equation}
where $q$ is the charge and $M$ is the mass parameter of the black hole. The corresponding single copy solution is 
\begin{equation}
	A_\m \dif x^\m= \phi(r) k_\m \dif x^\m =(8 G M + 2 G q^2 \log(r))\dif t\,, 
\end{equation}
which is just the Coulomb's solution with the identification $2Gq^2\rightarrow -Q$ where $Q$ is the charge of the point particle in Maxwell's theory.

Alternatively, one can consider the coupling to a free scalar as
\begin{equation}
    S= \int \dif^3 x\sqrt{-g}\left[\frac{1}{\k^2} R + \frac{1}{2} (\p \varphi)^2   \right]\,,
\end{equation}
with again a ghost sign for the matter term. Taking $\phi = \phi(r)$, a static black hole solution is obtained provided that $\varphi = p\,\theta $ ($p$: constant). The solution is given by
\begin{equation}
	\phi(r)=-2GM\log(r)\,,\quad \varphi=\sqrt{\frac{M}{2 \pi}}\theta\,.
\end{equation}
The single copy gauge field
\begin{equation}
    A_\m \dif x^\m= 2 G M \log(r) \dif t,
\end{equation}  
is again the Coulomb's solution this time with the identification $2 G M\rightarrow -Q$.   

The line element for the metric given in KS coordinates \eqref{KS} 
\begin{align}\label{KSline}
	\dif s^2 &= \e_{\m\n} \dif x^\m \dif x^\n + \f(r) (k_\m \dif x^\m)^2\nn\\
	&= -\left[1-\f(r)\right] \dif t^2 + \left[1+\f(r)\right]  \dif r^2 +2\f(r) \dif t \dif r+r^2\dif \t^2,
\end{align}
can be written in the Boyer-Lindsquit (BL) coordinates by the following coordinate transformation
\begin{equation}
	\dif t\rightarrow \dif t +\frac{\phi(r)}{1-\phi(r)} \dif r,
\end{equation}
as follows:
\begin{equation}\label{ds2}
	\dif s^2 = -h(r) \dif t^2 +\frac{1}{h(r)} \dif r^2 + r^2 \dif \theta^2\,,\quad h(r)=1-\phi(r).
\end{equation}
In BL coordinates, the existence of stable orbits can be easily studied. For timelike particles, the geodesic motion is governed by the equation
\begin{equation}
	\frac{1}{2} \mathcal{E}^2= \frac{1}{2}\left( \frac{\dif r}{\dif t} \right)^2 + V_{\text{eff}}\,,
\end{equation} 
where the effective potential is given by 
\begin{equation}
	V_{\text{eff}}=\frac{1}{2}\left(\frac{L^2}{r^2}+1\right)h(r)\,.
\end{equation}
The energy and the angular momentum of the particle are expressed in terms of the timelike and angular Killing vectors $\xi_{(t,\theta)}$ as
\begin{equation}
	\mathcal{E}=-g_{\m\n}\xi^\m_{(t)} u^\n\,,\qquad \qquad L=g_{\m\n}\xi^\m_{(\t)} u^\n\,,
\end{equation} 
where $u_\m$ is the velocity of the particle. The Newtonian potential $V_{\text{Newton}}$ can be obtained by neglecting $GL^2$ terms in the effective potential $V_{\text{eff}}$. The stable orbits were shown to exist in \cite{Alkac:2021seh} for the vector coupling and in \cite{CarrilloGonzalez:2019gof} for the scalar coupling.

When the single copy is the Coulomb's solution, both the double copy and the single copy has a singularity at $r=0$. In 3D, there are only three independent curvature invariants that can be constructed from contractions of the metric and the Riemann tensor \cite{Paulos:2010ke,Bueno:2022lhf,Gurses:2011fv}. For a general static solution with the line element \eqref{KSline}, they are given by
\begin{align}
	&R=\frac{\phi^\prime}{r}+\phi^{\prime\prime}, \label{R} \\
	&R^\m_{\ \n} R^\n_{\ \m}= \left(\frac{\phi^\pr}{r}\right)^2+\frac{1}{2}R^2,\\
	&R^\m_{\ \n} R^\n_{\ \r} R^\r_{\ \m}=\left(\frac{\phi^\pr}{r}\right)^3+\frac{1}{4}R^3,
\end{align}
and the electric field corresponding to single copy is
\begin{equation}\label{E}
	E(r)\equiv F_{rt}=-\phi^\prime(r)=h^\prime(r).
\end{equation}
We see that if $\dfrac{\phi^\prime}{r}$ and $\phi^{\prime\prime}$ are regular, then the double copy solution is regular and it can be checked by looking at the curvature scalar $R$ alone. However; it is not guaranteed by the regularity of the single copy electric field. For example, if one takes the scalar potential corresponding to a point charge in Born-Infeld electromagnetism
\begin{equation}\label{KSmetricfuncBI}
	\phi(r)=-Q\log\left[\frac{r+\sqrt{r^2+\dfrac{Q^2}{b^2}}}{2}\right],
\end{equation}
and studies a generalized static solution by assuming that the KS ansatz \eqref{KS} gets no correction, as suggested by \cite{Pasarin:2020qoa}, the curvature scalar and the single copy electric field read 
\begin{align}
	R(r)&= -\frac{Q^3}{r\left(r^2+\dfrac{Q^2}{r^2}\right)^{\nicefrac{3}{2}}}\nn\\
	E(r)&=\frac{Q}{\sqrt{r^2+\dfrac{Q^2}{b^2}}}\,.
\end{align}
Around $r=0$, one has
\begin{align}
	R(r)&=-\frac{b}{r}+\frac{3b^3r}{2Q^2}-\frac{15b^5r^3}{8Q^4}+\cO(r^5)\,,\nn \\
	E(r)&=b-\frac{b^3r^2}{2Q^2}+\frac{3b^5r^4}{8Q^4}+\cO(r^5)\,.
\end{align} 
This is a simple example where we explicitly see that one might have regular single copy electric fields despite having a singularity on the gravity side. If we start from a general KS scalar $\phi(r)$ that is regular at $r=0$ as follows,
\begin{equation}
	\phi(r)=a_0 + a_1 r+a_2 r^2 + \cO(r^3),
\end{equation}
where $a_i$'s ($i=0,1,2$) are arbitrary constants, we obtain
\begin{align}
	R(r)&=\frac{a_1}{r}+4a_2+9a_3 r+\cO(r^2),\\
	E(r)&=a_1+2a_2r+\cO(r^2).
\end{align}
Therefore; the necessary and sufficient condition for the regularity of both single and double copy solutions is $a_1=0$, which is the vanishing of the single copy electric field at the origin\footnote{In 4D, one also need to have $a_0=0$, which can be achieved by changing a different integration constant in obtaining the solution. However; this chances the asymptotic behaviour of the metric (see \cite{Pasarin:2020qoa} for details).}, i.e., $E(r=0)=0$.
In the next sections, by considering more generalized matter couplings, we will provide examples of different possibilities, in the usual context of KS double copy without making any assumptions such as made in \cite{Pasarin:2020qoa}.

\section{Einstein-Born-Infeld Theory}\label{sec:EBI}
One of the simplest and most natural generalizations of Einstein-Maxwell theory is Einstein-Born-Infeld theory described by the Lagrangian
\begin{equation}\label{ebi-action}
S=\int \mathrm{d}^{3} x \sqrt{-g}\left[\frac{\zeta_{1}}{\kappa^{2}}R+\zeta_{2} \cL_{\text{BI}}(f) \right], \quad \kappa^{2}=8 \pi G\,,
\end{equation}
where we have introduced $\z_i=\pm1$ $(i=1,2)$ to control the sign of the kinetic terms ($-1$: ghost, $+1$: not ghost) and the Lagrangian of the Born-Infeld electrodynamics is given by \cite{BI}
\begin{equation}\label{BI-Lag}
\cL_{\text{BI}}(f)=\frac{b^2}{2\pi} \left( 1-\sqrt{1+\frac{ f^2}{2 b^2}} \right),\qquad f_{\m\n}=2\p_{\left[\m\right.} a_{\left.\n \right]},
\end{equation}
which reduces to that of Maxwell theory as $b \goesto \infty$. Assuming a static line element of the KS form \eqref{KSline} and $a_\m \dif x^\m = a_t(r)\dif t$, the matter equations
\begin{equation}
\partial_\m \left(\dfrac{\sqrt{-g}f^{\m \n}}{\sqrt{1+\dfrac{ f^2}{2 b^2}} } \right)=0\,,
\end{equation}
are solved by the following scalar potential and the corresponding independent nonzero component of the field strength tensor
\begin{equation}
	a_{t}(r)=-q\log\left[\frac{r+\psi(r)}{2}\right],\qquad f_{rt}=\frac{q}{\psi(r)},
\end{equation}
where
\begin{equation}
\psi(r)=\sqrt{r^2+\dfrac{q^2}{b^2}}.
\end{equation}
With the energy-momentum tensor
\begin{equation}\label{Engmom-BI}
T_{\m\n}=\frac{1}{2 \pi}\frac{f_\m ^{\ \a}f_{\n\a}}{\sqrt{1+\dfrac{ f^2}{2 b^2}}}+g_{\m\n}\cL_{\text{BI}}(f),
\end{equation}
the trace-reversed Einstein equations read
\begin{equation}
R_{\mu \n}-2 \Lambda g_{\mu \n}=\zeta \frac{\kappa^{2}}{4 \pi}\left( \frac{2 f_{\mu}^{\ \alpha} f_{v \alpha}-g_{\mu \n} f^2}{\sqrt{1+\dfrac{ f^2}{2 b^2}}} + 2\, g_{\m\n} \cL_{\text{BI}}(f)\right),
\end{equation}
where $\z=\zeta_{1}\zeta_{2}$. The independent components of the left-hand side of the equations are
\begin{align}
&(\mathrm{LHS})_{t t}=\frac{[\phi(r)-1]\left[r \phi^{\prime \prime}(r)+\phi^{\prime}(r)-4 \Lambda r\right]}{2 r}, \nn\\
&(\mathrm{LHS})_{t r}=\frac{\phi(r)\left[r \phi^{\prime \prime}(r)+\phi^{\prime}(r)-4 \Lambda r\right]}{2 r}, \nn\\
&(\mathrm{LHS})_{r r}=\frac{[\phi(r)+1]\left[r \phi^{\prime \prime}(r)+\phi^{\prime}(r)-4 \Lambda r\right]}{2 r}, \nn\\
&(\mathrm{LHS})_{\theta \theta}=r \phi^{\prime}(r)-2 \Lambda r^{2}\,,
\end{align}
with the following components at the right-hand side
\begin{align}
&(\mathrm{RHS})_{t t}=\frac{2 b^2 \zeta  G \left[\phi (r)-1\right] \left[r-\psi(r)\right]^2}{r \psi(r)}, \nn \\
&(\mathrm{RHS})_{t r}=-\frac{2 b^2 \zeta  G \phi (r) \left[r-\psi(r)\right]^2}{r \psi(r)},\nn \\
&(\mathrm{RHS})_{r r}=\frac{2 b^2 \zeta  G \left[1+\phi (r)\right] \left[r-\psi(r)\right]^2}{r \psi(r)},\nn \\
&(\mathrm{RHS})_{\theta \theta}= -4 b^2 \zeta  G r \left[r-\psi(r)\right].
\end{align}
Similar to the Einstein-Maxwell case, the $\t\t$-component is the easiest one to solve and it yields
\begin{equation}\label{BImetricfunc}
\phi(r)=  - 8GM+\z G q^2+2\z Gb^2\left[r^2-r\psi(r)-\frac{q^2}{b^2}\log\left(\frac{r+\psi(r)}{2}\right)\right],
\end{equation}
which also solves the other components\footnote{The static solution for a nonzero cosmological constant and no ghost sign in the action was given in \cite{Cataldo:1999wr,Myung:2008kd}.}. Note that we have chosen the integration constant such that the expansion around $b \goesto \infty$
\begin{equation}
	\phi(r)=-8GM+2\z Gq^2\log(r)+\frac{\z G q^4}{4b^2r^2}+\cO\left[\frac{1}{b^3}\right],
\end{equation}
gives the result for $\z=-1$ in the EM case \eqref{maxsol} at the leading order. The gravitational field in the Newtonian limit is given by
\begin{align}
\vec{g}&=\frac{1}{2}\vec{\nabla}\phi=-\frac{1}{2}\vec{\nabla}h,\nn\\
&=\frac{2\z G \left[q^2+b^2r\left(r-\sqrt{r^2-\dfrac{q^2}{b^2}}\right)\right]}{\psi(r)} \hat{r},
\end{align}
which is attractive everywhere when $\z=-1$. Therefore, we again need to choose one ghost sign in the action. For $\z=-1$, the Ricci scalar and the electric field corresponding to the single copy $A_\m=\phi k_\m$ are given by
\begin{align}
	R(r)&=-\frac{4G\left[q^2+2b^2r\left(r-\psi(r)\right)\right]}{r\psi(r)},\nn\\
	E(r)&=\frac{4G\left[q^2+b^2r\left(r-\psi(r)\right)\right]}{\psi(r)}.
\end{align}
Checking their behavior as $r \goesto 0$, 
\begin{align}
	R(r)&=-\frac{4Gbq}{r}+8Gb^2-\frac{6Gb^3r}{q}+\cO(r^3),\nn\\
	E(r)&=4Gbq-4Gb^2r+\frac{2Gb^3r^2}{q}+\cO(r^3),	
\end{align}
one sees that, while the single copy electric field is regular around the origin, we have a singularity on the gravity side. 

\begin{figure}
	\centering
	\includegraphics[width=0.95\linewidth]{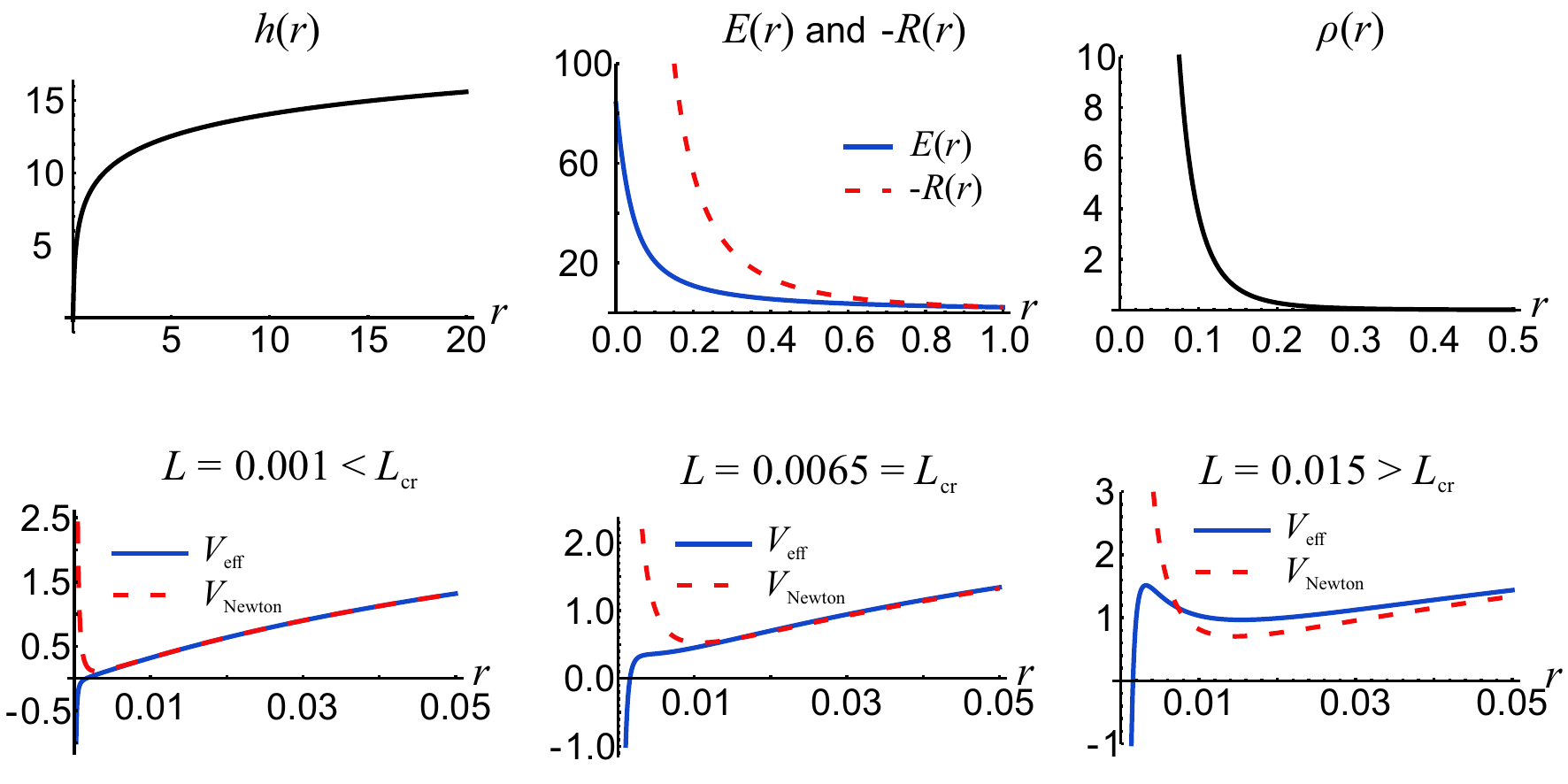}
	\caption{Details of the static solution of Einstein-Born-Infeld theory for $(G=1, M=1, q=0.525, b=10)$.}
	\label{fig:EBI}
\end{figure}

For particle orbits, one can ensure to preserve the following main properties of the static solution of Einstein-Maxwell theory by choosing an appropriate set of parameters: i) The Newtonian potential $V_{\text{Newton}}$ possesses an infinite barrier at short distances and becomes equal to the effective potential $V_{\text{eff}}$ at large distances. ii) Timelike particles are forbidden to reach the infinity due to the logaritmic behaviour of the potential as $r \goesto \infty$. iii) There is a critical value $L_c$ of the angular momentum of the particle such that, when $L>L_c$, the effective potential $V_{\text{eff}}$ develops a local minimum and a local maximum, making stable orbits possible. We refer the reader to Figure \ref{fig:EBI} for an explicit demonstration of these properties and the regularity of the solutions, together with the charge density in the gauge theory.

Before proceeding further, we would like to note that, similar to the Einstein-Maxwell theory, one can again make use of the duality of scalars and gauge vectors to realize a solution with the same physical properties. The matter part of the Lagrangian is given by
\begin{equation}
\cL_{\text{scalar}}(\varphi) = -\frac{b^2}{2\pi} \left(1- \sqrt{1+\frac{2\pi}{b^2} (\p \varphi)^2} \right),
\end{equation}
which reduces to the free scalar Lagrangian as $b\goesto\infty$ and the scalar field should read
\begin{equation}
\varphi=\sqrt{\dfrac{M}{2\pi}}\theta, \qquad \qquad M=q^2,
\end{equation}
in order to obtain the same solution as that of Einstein-Born-Infeld theory up to an integration constant.
\section{Generalization of the Scalar Coupling}\label{sec:sca}
In this section, we will study a generalization of the scalar coupling which not only allows different possibilities regarding the regularity of the single and double copies, but also demonstrates a simple relation between horizons on the gravity side and the corresponding electric field in Maxwell's theory, which is based on the following observation \cite{Easson:2020esh}: Since the single copy electric field is equal to the derivative of the metric function in BL coordinates [$E(r)=h^\prime(r)$], it becomes zero at a maximum, a minimum or a saddle point. Since the horizons are located at the zeros of the metric function [$h(r_i)=0$], there should exist at least one point between two adjacent horizons where the electric field is zero, corresponding to a minimum or maximum. The theory that we will consider is described by the action \cite{Bueno:2021krl}
\begin{equation}
S=\int \dif^3 x \sqrt{-g}\left[\frac{1}{\k^2}(R-2\L) -Q \right], 
\end{equation}
where the matter term is given by
\begin{align}
	Q =&  \sum_{n=1} \alpha_{n} \ell^{2(n-1)}(\partial \vf)^{2 n}\nn\\
	&-\sum_{m=0} \beta_{m} \ell^{2(m+1)}(\partial \vf)^{2 m} \left[(3+2 m) R^{\m\n} \partial_{\m} \vf \partial_{\n} \vf- R(\partial \vf)^{2}\right].
\end{align}
Here, $\a_n$ and $\b_m$ are arbitrary dimensionless constants. The trace-reversed Einstein equations are in the following form:
\begin{equation}
	R_{\m\n} = \frac{\k^2}{2} \T_{\m\n}, \qquad \qquad \T_{\m\n}=\frac{1}{\Gamma(\vf)}\left[\tT_{\m\n}+\frac{4\L}{\k^2}g_{\m\n}\right].\label{grav}
\end{equation}
We give the expressions for $\Gamma(\vf), \tT_{\m\n}$ and the field equation for the scalar field in Appendix \ref{sec:app} since they are quite cumbersome and do not play a direct role in our discussion. Applying the usual prescription, we again obtain Maxwell's equations with a source defined in terms of the $\T$-tensor as follows:
\begin{equation}
	\p_\n F^{\n\m} = g J^\m, \qquad J^\m = 4\,\T^{\m}_{\ 0}.
\end{equation}
This theory admits a family of black holes and horizonless spacetimes whose line elements in BL coordinates are in the form \eqref{ds2}. (see \cite{Bueno:2021krl} for the most general form of the solution). For our purposes, it is enough to take non-zero values for $(\a_1,\a_2,\a_3,\b_0,\b_1)$ and set all the other constants to zero. In this case, for $\vf=p\, \t$ ($p$:constant), all the field equations are solved if the metric function is given by
\begin{equation}
	h(r)=\frac{(1+8GM) r^4-\Lambda  r^6-8 \pi G \a_1  p^2 r^4 \log (r)+4 \pi G \a_2 \ell^2  p^4+2 \pi G \a_3 \ell^4  p^6 r^2}{8 \pi G  \b_0  \ell^2 p^2 r^2+24 \pi G  \b_1  \ell^4 p^4+r^4},
\end{equation}
where we have chosen the integration constant such that such that we recover the free scalar case when $p=\sqrt{\dfrac{M}{2\pi}}$, $\a_1 = -\dfrac{1}{2}$, $\a_2=\a_3=\b_1=\b_2=\L=0$. The curvature scalar and the single copy electric field can be easily calculated from equations \eqref{R} and \eqref{E} respectively. Their expansions around $r=0$ are given by
\begin{align}
	R(r)&=\frac{\a_3 \b_0-6 \a_2 \b_1}{6 \b_1^2 \ell^2}+\frac{\left[c_1+c_2 \log(r)\right]r^2}{216\pi G\b_1^3 \ell^4 p^4}+\cO(r^3)\,\nn\\
	E(r)&=-\frac{r \left(\a_3 \b_0-6 \a_2 \b_1\right)}{18 \b_1^2 \ell^2}+\cO\left(r^3\right),
\end{align}
where $c_1=c_1(G,M,p,\a_2,\a_3,\b_0,\b_1)$ and $c_2=c_2(G,p,\a_1,\b_1)$ are constants. From the expansions, we see that one can obtain non-singular single and double copies by taking $\b_1 \neq 0$. When $\b_0=\b_1=0$, the expansions around $r=0$ become
\begin{align}
	R(r)&=-\frac{24\pi G \a_3 \ell^4 p^6}{r^6}-\frac{8\pi G \a_2 \ell^2 p^4}{r^4}+\frac{8 \pi G \a_1 p^2}{r^2} + 6 \L + \cO(r^3),\nn\\
	E(r)&= -\frac{8 \pi G \a_3 \ell^4 p^6}{r^5}-\frac{8 \pi G \a_2 \ell^2 p^4}{r^3}-\frac{8 \pi G \a_1 p^2}{r} - 2 \L r + \cO(r^3).
\end{align}
The expression for the electric field shows that one has a point charge at the origin with $Q=-8 \pi G \a_1 p^2$.

We present four cases by using different sets of parameters, which are given in Table \ref{table}:
\begin{itemize}
	\item \underline{Case I:}\,Taking a nonzero value for $\b_1$, we obtain regular single and double copies. This is a horizonless geometry and the electric field $E(r)$ becomes zero at two points, which are maximum and minimum of the metric function $h(r)$. 
	
	\item \underline{Case II:}\,Again taking $\b_1 \neq 0$ guarantees the regularity of the single and double copies. There is one event horizon and the electric field $E(r)$ is zero only at the origin, where the minimum of the metric function $h(r)$ occurs. Stable orbits exist when $L>L_{\text{cr}}$.

	\item \underline{Case III:} $\b_1 \neq 0$ yields regular single and double copies. We have two event horizons and the electic field becomes zero at two points: A local maximum ($r=0$) and a global minimum located between two horizons.
	
	\item\underline{Case IV:} $\b_1=0$ gives single and double copies which are singular at the origin. There are three event horizons and the electric field is zero at the following points: a local maximum between the first and the second horizons, and a local mininum between the second and the third horizons.
\end{itemize}

\begin{figure}
	\centering
	\includegraphics[width=0.92\linewidth]{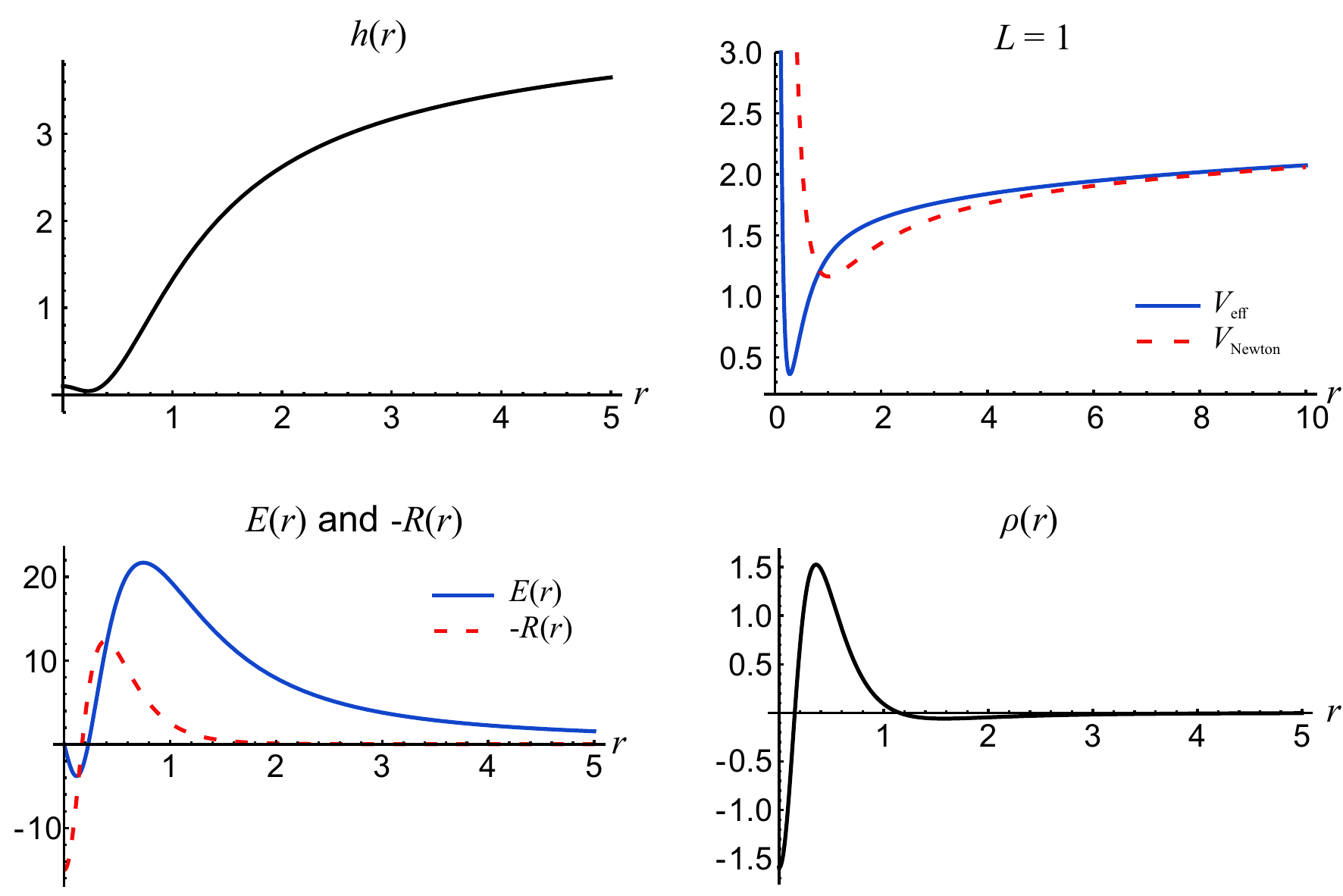}
	\caption{Details of Case I presented in Section \ref{sec:sca} for parameters given in Table \ref{table}.\label{fig:case1}}
\end{figure}
\begin{figure}
	\centering
	\includegraphics[width=0.95\linewidth]{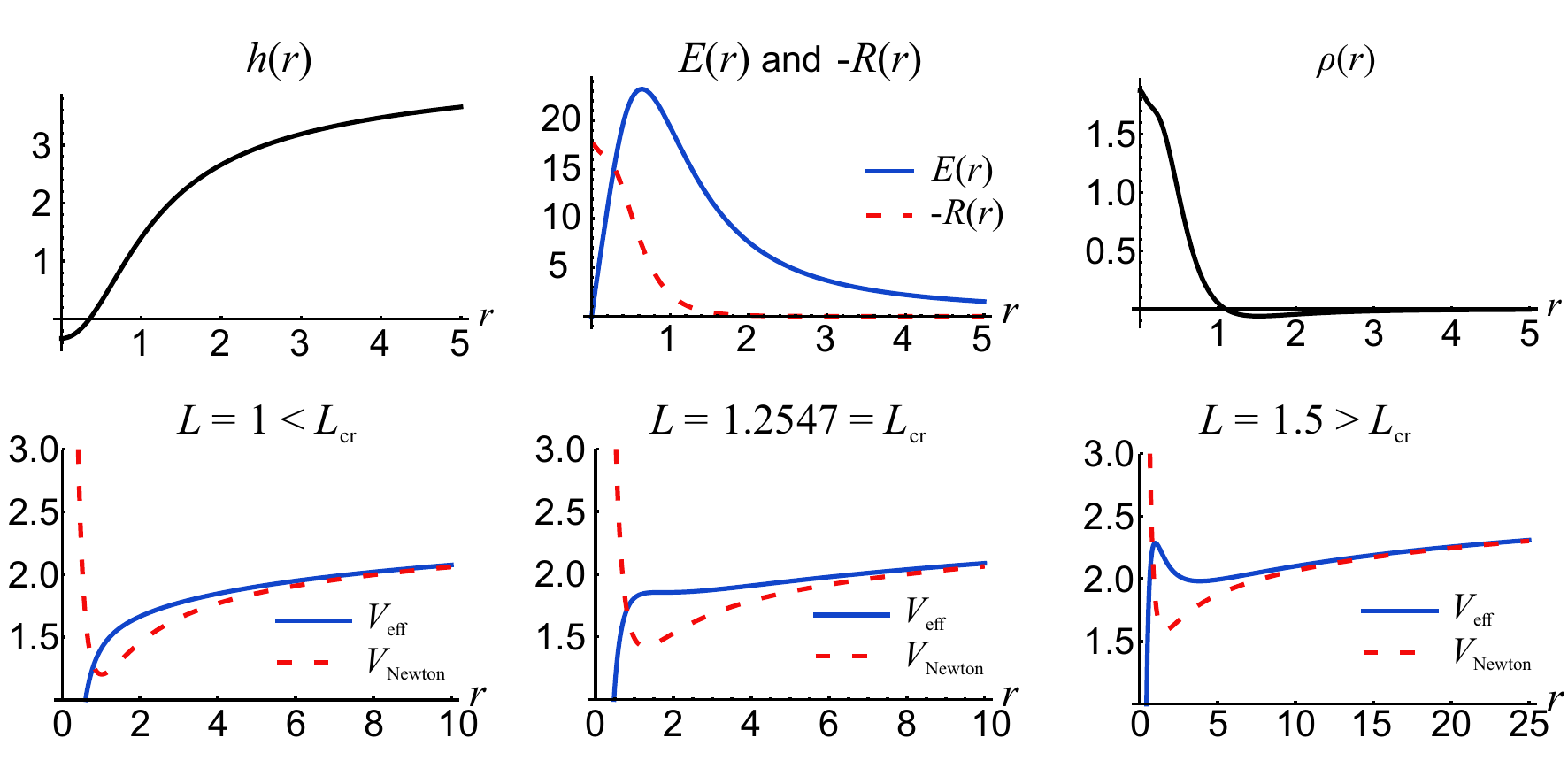}
	\caption{Details of Case II presented in Section \ref{sec:sca} for parameters given in Table \ref{table}.\label{fig:case2}}
\end{figure}
\begin{figure}
	\centering
	\includegraphics[width=0.92\linewidth]{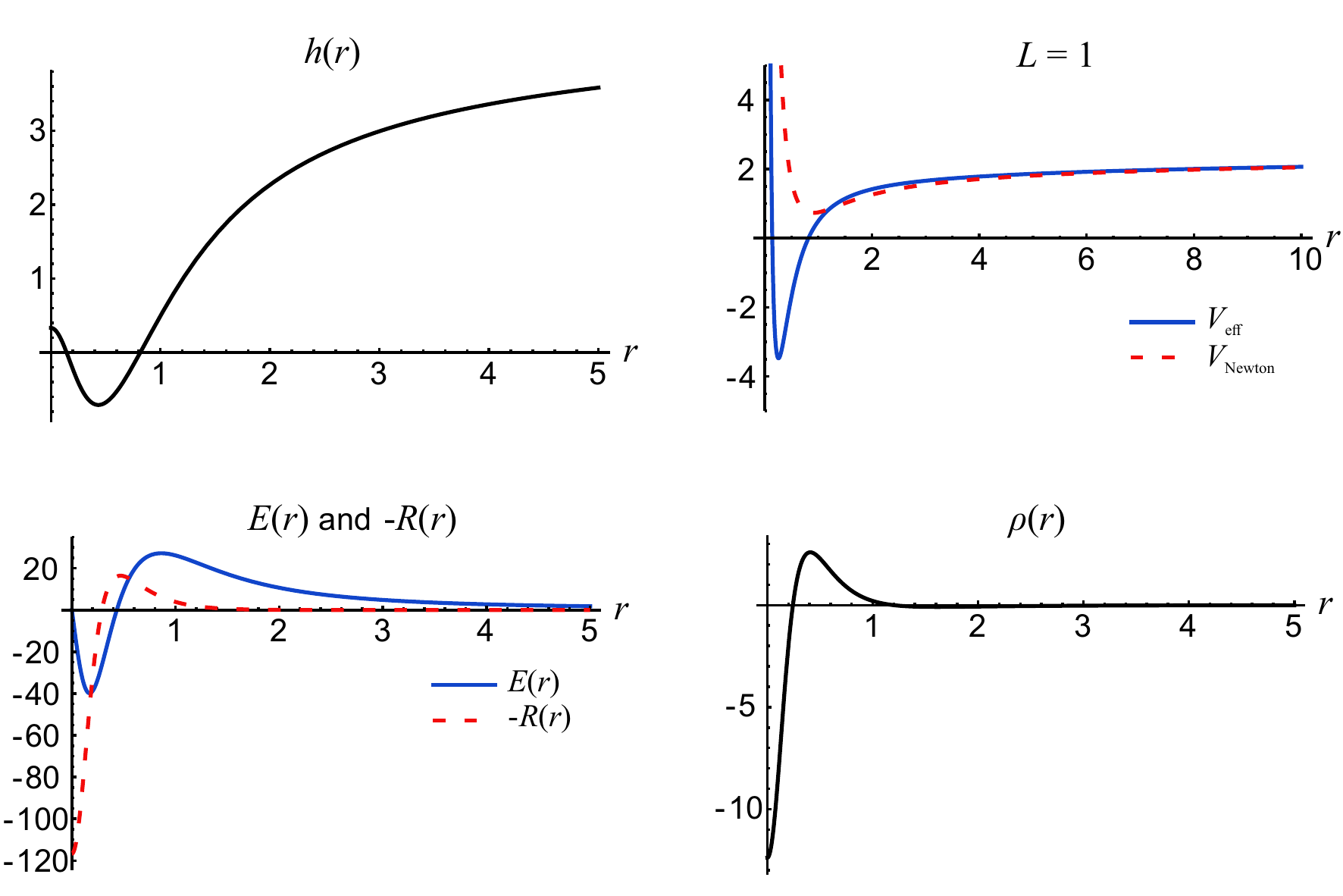}
	\caption{Details of Case III presented in Section \ref{sec:sca} for parameters given in Table \ref{table}.\label{fig:case3}}
\end{figure}
\begin{figure}
	\centering
	\includegraphics[width=0.92\linewidth]{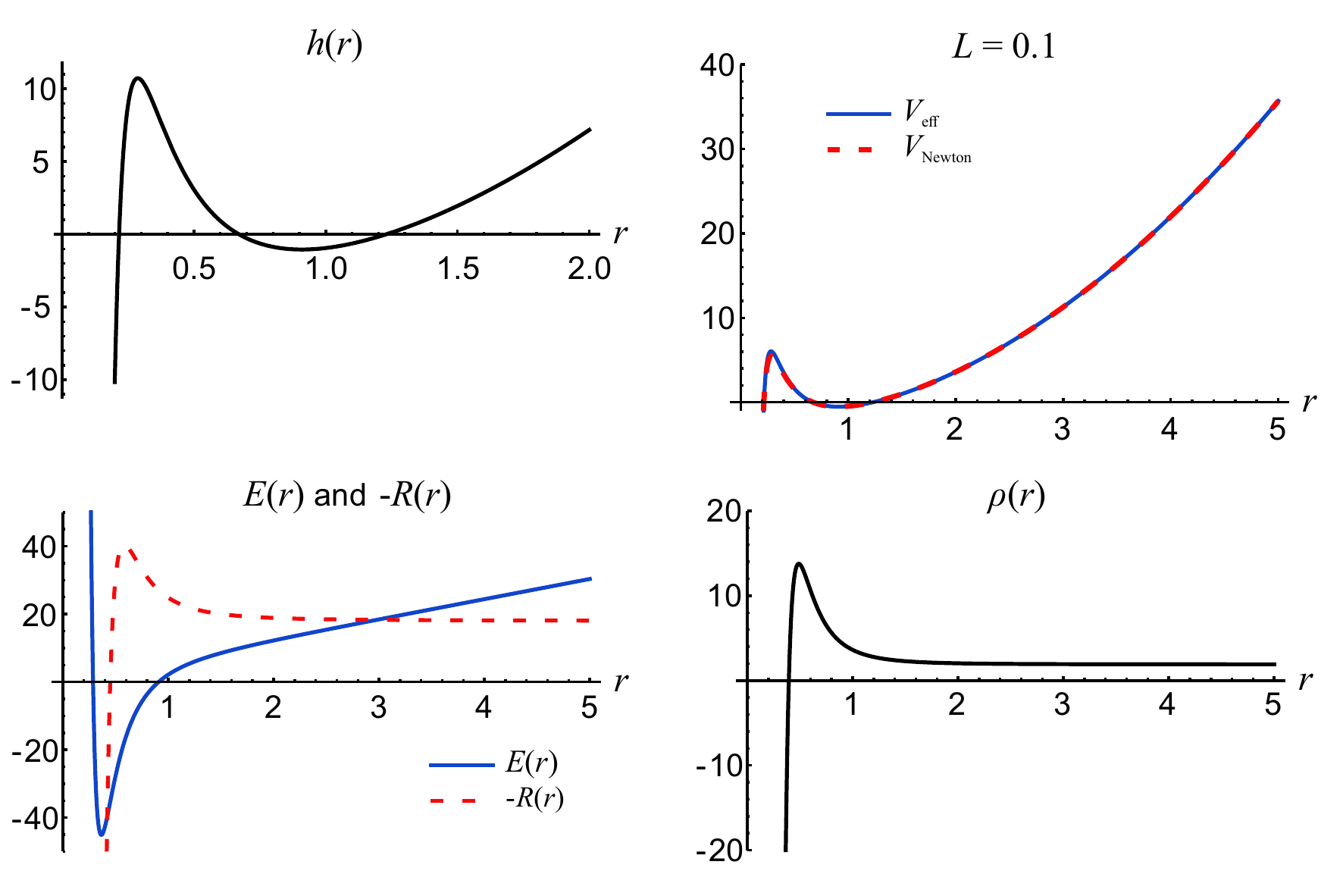}
	\caption{Details of Case IV presented in Section \ref{sec:sca} for parameters given in Table \ref{table}. Note that in addition to the nonlocal charge distribution shown in the figure, one has a point charge with $Q=2$ at the origin.\label{fig:case4}}
\end{figure}

All the details can be seen in Figures \ref{fig:case1}, \ref{fig:case2}, \ref{fig:case3} and \ref{fig:case4}. Although, stable orbits exist for all values of the angular momentum $L$ in cases I, III and IV; there is a critical value $L_{\text{cr}}$, beyond which, there arises a second region where a particle in a stable orbit can be present. We do not show them explicitly since it is sufficient to show the existence of one such region for our purposes. As we have shown at the end of Section \ref{sec:KS} on general grounds, the necessary and sufficient condition for the regularity of the solutions on both gravity and gauge theory sides is the vanishing of the electric field at the origin, which is realized in Cases II and III and can be explicitly seen in Figures \ref{fig:case2},\ref{fig:case3}.
\begin{table}
	\centering
	\begin{tabular}{|c|c|c|c|c|c|c|c|c|c|c|} 
		\hline &&&&&&&&&& \\ [-13pt]
		& $G$ & $M$ & $p$ & $\ell$ & $\L$& $\a_1$& $\a_2$& $\a_3$ & $\b_0$ & $\b_1$ \\
		\hline  &&&&&&&&&& \\ [-13pt]
		Case I & \multirow{2}{*}{$1$} & \multirow{2}{*}{$\dfrac{1}{4}$} & \multirow{2}{*}{$\sqrt{\dfrac{M}{2\pi}}$} & \multirow{2}{*}{$1$} & \multirow{2}{*}{$0$} & \multirow{2}{*}{$-\dfrac{1}{2}$} & \multirow{2}{*}{$-10$} & \multirow{2}{*}{$30$}& \multirow{2}{*}{$1$}& \multirow{2}{*}{$1$} \\ (No Horizon)&&&&&&&&&& \\
		\hline &&&&&&&&&& \\ [-13pt]
		Case II & \multirow{2}{*}{$1$} & \multirow{2}{*}{$\dfrac{1}{4}$} & \multirow{2}{*}{$\sqrt{\dfrac{M}{2\pi}}$} & \multirow{2}{*}{$1$} & \multirow{2}{*}{$0$} & \multirow{2}{*}{$-\dfrac{1}{2}$} & \multirow{2}{*}{$1$} & \multirow{2}{*}{$-5$}& \multirow{2}{*}{$1$}& \multirow{2}{*}{$1$} \\ (Single Horizon)&&&&&&&&&& \\ [2pt]
		\hline&&&&&&&&&& \\ [-13pt]
		Case III & \multirow{2}{*}{$1$} & \multirow{2}{*}{$\dfrac{1}{4}$} & \multirow{2}{*}{$\sqrt{\dfrac{M}{2\pi}}$} & \multirow{2}{*}{$1$} & \multirow{2}{*}{$0$} & \multirow{2}{*}{$-\dfrac{1}{2}$} & \multirow{2}{*}{$-100$} & \multirow{2}{*}{$100$}& \multirow{2}{*}{$1$}& \multirow{2}{*}{$1$} \\ (Two Horizons)&&&&&&&&&& \\ [2pt]
		\hline &&&&&&&&&& \\ [-13pt]
		Case IV & \multirow{2}{*}{$1$} & \multirow{2}{*}{$1$} & \multirow{2}{*}{$\sqrt{\dfrac{M}{2\pi}}$} & \multirow{2}{*}{$1$} & \multirow{2}{*}{$-3$} & \multirow{2}{*}{$-\dfrac{1}{2}$} & \multirow{2}{*}{$10$} & \multirow{2}{*}{$-5$}& \multirow{2}{*}{$0$}& \multirow{2}{*}{$0$} \\ (Three Horizons)&&&&&&&&&& \\ [2pt]
		\hline
	\end{tabular}\caption{Choice of parameters for different solutions of the scalar theory. We always take $\sqrt{\frac{M}{2\pi}}$, $\a_1=-\frac{1}{2}$, as in the free scalar case, and $G=1$, which leads to $\p_\n F^{\n\m} = 2\pi J^\m$.\label{table}}
\end{table}
\section{Summary and Discussions}\label{sec:sum}
In this paper, we have studied generalizations of the matter couplings in 3D admitting a static black hole solution which gives rise to the Coulomb's solution as its single copy. For these matter couplings, Einstein-Maxwell theory or GR minimally coupled to a free scalar field, both the double copy and the single copy solution has a singularity at the origin. As a generalization of the former, we studied Einstein-Born-Infeld theory and showed that although the static black hole solution, which admits stable particle orbits, is singular, the single copy electric field is regular at the origin. For the latter, we have investigated a theory recently discovered in \cite{Bueno:2021krl}, which forms an extremely useful theoretical laboratory since the most general solution offers different possibilities regarding the regularity of the black hole solution and the number of event horizons. We have given examples where both the double and single copy are regular. Morever, starting from a horizonless geometry, we have considered spacetimes with increasing number of event horizons and exhibited the relation between the event horizons on the gravity side and the corresponding electric field in Maxwell's theory. 

All these examples show that many physically important properties of the KS double copy can also be realized in 3D with the most notable exception that, in the simplest case where the single copy is the Coulomb solution, the double copy is a non-vacuum solution which can be obtained by taking the Einstein-Hilbert term or the matter term with a ghost sign in the action. In the generalizations that we have considered in this paper, we introduced the couplings such that one recovers the Coulomb case in an appropriate limit. However; a wide range of different possibilities exists without this requirement.

As a final note, we would like to mention that there exists a different interpretation of the single copy in the case of a non-minimal coupling on the gravity side. Writing the gravitational field equations \eqref{grav} by introducing an effective Newton constant as
\begin{equation}
	R_{\m\n} = \frac{\k_{\text{eff}}^2}{2} \T^\pr_{\m\n}, \qquad \qquad \k_{\text{eff}}^2=\frac{\k^2}{\Gamma(\vf)},
\end{equation}
where
\begin{equation}
	\T^\pr_{\m\n} = \tT_{\m\n}+\frac{4\L}{\k^2},
\end{equation}
one can obtain solutions to Maxwell's equations with an effective gauge coupling
\begin{equation}
	\p_{\n} F^{\n\m} = g_{\text{eff}} J^\m, \qquad \qquad J^\m = 4 \T^{\pr \m}_{\ 0},
\end{equation} 
with the identification $\k^2_{\text{eff}} \goesto 4g_{\text{eff}}$.
In this picture, the cosmological constant plays its usual role in the case of a minimal matter coupling and produces a constant charge density filling all space. However; a dynamical mechanism for the evolution of the gauge coupling seems to be missing on the gauge theory side. This might be an interesting direction for future study. 
\section*{Acknowledgments}
	We thank Hasan Y{\i}lmaz for his help in creating high-resolution figures and the anonymous referee for useful suggestions.
\appendix
\section{Details of the Field Equations of the Scalar Theory}\label{sec:app}
\begin{equation}
	\Gamma(\vf)=1-\dfrac{\k^2}{2}\sum_{m=0}\b_m \ell^{2(m+1)} (\partial \vf)^{2(m+1)}
\end{equation}
\begin{equation}
	\tilde{T}_{\m\n}=\tilde{T}_{\m\n}^{(1)}  +\tilde{T}_{\m\n}^{(2)} +\tilde{T}_{\m\n}^{(3)},
\end{equation}
where
\begin{align}
	\tilde{T}_{\m\n}^{(1)} =& \sum_{n=1} \alpha_{n}(\partial \vf)^{2(n-1)} \ell^{2(n-1)}\left(n \partial_{\m} \vf \partial_{\n} \vf- g_{\m \n}(n-1)(\partial \vf)^{2}\right)\nn\\
	\tilde{T}_{\m\n}^{(2)}=&\sum_{m=0} \beta_{m}\ell^{2(m+1)}(\partial \vf)^{2(m-1)} \left[2(3+2 m) R_{\a(\m} \partial_{\n)} \vf \partial^{\a} \vf(\partial \vf)^{2}+m(3+2 m) \partial_{\m} \vf \partial_{\n} \vf R^{\a \b} \partial_{\a} \vf \partial_{\b} \vf\right. \nn \\
	&\left.-(m+1) (\partial \vf)^{2}\left( \partial_{\m} \vf \partial_{\n} \vf R+ g_{\m\n}  R^{\a \b} \partial_{\a} \vf \partial_{\b} \vf - g_{\m\n}R (\partial  \vf)^{2}\right) \right],\nn\\
	\tilde{T}_{\m\n}^{(3)}=&\,\nabla^{\a} \nabla_{(\m} E_{\n) \a}-\frac{1}{2} \Box E_{\m \n}-\frac{1}{2} g_{\m \n} \Box E, 
\end{align}
with
\begin{equation}
E_{\m \n}=\sum_{m=0} \beta_{m} \ell^{2(m+1)}(\partial \vf)^{2 m}\left[(3+2 m) \partial_{\m} \vf \partial_{\n} \vf-g_{\m \n}(\partial \vf)^{2}\right].
\end{equation}
The equation for the scalar field reads
\begin{align}
0&=2 \nabla_{\m}\left[\sum_{n=1} n \alpha_{n} \ell^{2 n-1}(\partial \phi)^{2(n-1)} \partial^{\m} \phi\right. \\
&\left.-\sum_{m=0} \beta_{m} \ell^{2(m+1)}(\partial \phi)^{2(m-1)}\left[m(3+2 m) \partial^{\m} \phi R^{\a \b} \partial_{\a} \phi \partial_{\b} \phi+(3+2 m)(\partial \phi)^{2} R^{\m \a} \partial_{\a} \phi-(m+1) R(\partial \phi)^{2} \partial^{\m} \phi\right]\right].
\end{align}

\singlespacing

\end{document}